\documentclass[12pt]{article}
\usepackage[colorlinks=true, allcolors=blue]{hyperref}
\usepackage[colorinlistoftodos, textsize=tiny]{todonotes}
\usepackage{bm}
\usepackage{subcaption}
\usepackage{diagbox}
\usepackage{times}

\newcommand{\aj}{Astron. J.}   
\newcommand{\apj}{Astrophys. J.}   
\newcommand{\apjs}{Astrophys. J. Suppl. Ser.}   
\newcommand{\apss}{Astrophys. Space Sci.}   
\newcommand{\aap}{Astron. Astrophys.}   
\newcommand{\mnras}{Mon. Not. R. Astron. Soc.}   
\newcommand{\nastro}{Nat. Astron.} 
\newcommand{\na}{New Astron.}   
\newcommand{\pasp}{Publ. Astron. Soc. Pac.}   
\newcommand{\raa}{Res. Astron. Astrophys.} 

\usepackage[english]{babel}
\usepackage[T1]{fontenc}
\usepackage{booktabs} 

\usepackage{amsmath,amsfonts,amssymb}
\usepackage{graphicx}
\usepackage{threeparttable}

\usepackage{bm}
\usepackage{subcaption}

\usepackage{color}
\usepackage{pdfcomment}

\usepackage{amsmath}
\usepackage{lipsum}
\usepackage{graphicx}
\usepackage{amssymb}
\usepackage{color}
\usepackage{cite}
\usepackage{xcolor} 

\newcommand{\ks}{K_{\rm s}}
\newcommand{\ds}{\dot{s}}
\newcommand{\da}{\dot{\alpha}}
\newcommand{\ssa}{s,\dot{s},\dot{\alpha}}
\newcommand{\sa}{\dot{s},\dot{\alpha}}

\usepackage[utf8x]{inputenc}

\presetkeys%
    {todonotes}%
    {noline,backgroundcolor=green!40}{}

\title{An Earth-Mass Planet and a Brown Dwarf in Orbit Around a White Dwarf}
\author{Keming Zhang$^{1,2,3}$, Weicheng Zang$^{4,5}$, Kareem El-Badry$^6$, Jessica~R.~Lu$^3$, \\Joshua S. Bloom$^3$, Eric Agol$^7$, B. Scott Gaudi$^8$, Quinn Konopacky$^1$,\\ Natalie LeBaron$^3$, Shude Mao$^4$, Sean Terry$^{3,9,10}$}
\date{\small{%
    $^1$\textit{Department of Astronomy and Astrophysics, University of California, San Diego, CA 92122, USA} \\
    $^2$\textit{Hal{\i}c{\i}o\u{g}lu Data Science Institute, University of California, San Diego, CA 92122, USA} \\
    $^3$\textit{Department of Astronomy, University of California, Berkeley, CA 94720, USA} \\
    $^4$\textit{Department of Astronomy, Tsinghua University, Beijing 100084, China}\\
    $^5$\textit{Center for Astrophysics | Harvard \& Smithsonian, Cambridge, MA 02138, USA} \\
    $^6$\textit{Department of Astronomy, California Institute of Technology, Pasadena, CA 91125, USA} \\
    $^7$\textit{Department of Astronomy, University of Washington, Seattle, WA 98195, USA} \\
    $^8$\textit{Department of Astronomy, The Ohio State University, Columbus, OH 43210, USA} \\
    $^9$\textit{Department of Astronomy, University of Maryland, College Park, MD 20742, USA}\\
    $^{10}$\textit{Code 667, NASA Goddard Space Flight Center, Greenbelt, MD, 20771, USA}}}

\begin{document}
\baselineskip12pt
\maketitle
{\bf
Terrestrial planets born beyond 1--3 AU have been theorized to avoid being engulfed during the red-giant phases of their host stars. Nevertheless, only a few gas-giant planets have been observed around white dwarfs (WDs) --- the end product left behind by a red giant.
Here we report on evidence that the lens system that produced the microlensing event KMT-2020-BLG-0414 is composed of a WD orbited by an Earth-mass planet and a brown dwarf (BD) companion, as shown by the non-detection of the lens flux using Keck Adaptive Optics (AO).
From microlensing orbital motion constraints, we determine the planet to be a $\mathbf{1.9\pm0.2}$ Earth-mass ($M_\oplus$) planet at a physical separation of $\mathbf{2.1\pm0.2}$ au from the WD during the event.
By considering the system evolutionary history, we determine the BD companion to have a projected separation of 22 au from the WD, and reject an alternative model that places the BD at 0.2 au.
Given planetary orbital expansion during the final evolutionary stages of the host star, this Earth-mass planet may have existed in an initial orbit close to 1 au, thereby offering 
a glimpse into the possible survival of planet Earth in the distant future.}

The ultra-high-magnification nature of the microlensing event KMT-2020-BLG-0414 \\(KB200414 hereafter) has previously prompted intensive photometric follow-up observations around the peak of the event on July 11, 2020.
Modeling of the densely-sampled light curve subsequently revealed a three-body lens system consisting of a low-mass-ratio planet ($q\sim10^{-5}$) and a brown dwarf companion orbiting a sub-solar-mass host star \cite{zang_earth-mass_2021}.
Owing to intrinsic microlensing degeneracies \cite{dominik_binary_1999,jiang_ogle-2003-blg-238_2004,poindexter_systematic_2005}, there exist four distinct models that explain the light-curve data equally well. Among the four models, the projected separation for the brown dwarf companion can be very close ($\sim$0.2 au) or very wide ($\sim$20 au), and the lens-source relative proper motion can be either in the north-east or south-east (NE/SE) directions, which is associated with distinct microlensing parallax constraints.
On the other hand, the planet properties are consistent across the four models, all of which indicate an approximately Earth-mass planet at a projected separation of around 1--2 au.

For KB200414, the mass of the primary lens star (Table 1) as inferred from the finite-source and microlensing-parallax effects indicates that it is either a main-sequence (MS) star or a WD stellar remnant.
A MS lens star is expected to have a similar apparent brightness to the microlensing source star, whose apparent brightness is known from the magnification profile.
On the other hand, a WD lens is expected to be fainter by 6--8 magnitudes, making it practically undetectable under the glare of the source star.
Therefore, the two scenarios could be distinguished by measuring the total brightness at the event location prior to, or long after the event.
OGLE-III pre-event imaging (Figure 1a) measured the total brightness at the event location to be $I_{\rm base}$ = 18.46 $\pm$ 0.09, which implies a total blended flux of $I\sim19.3$ on top of the unmagnified source star brightness of $I\sim19.1$.
This blended light was originally reported by ref.\cite{zang_earth-mass_2021} as consistent with the expected MS lens brightness (Table 1),
but may also be attributed nearby field stars that cannot be resolved with seeing limited imaging.

To further constrain the lens brightness, we observed the location of KB200414 in the K-short infrared pass-band ($\ks$; 2.146 $\mu$m) with laser-guide-star AO \cite{wizinowich_w_2006,van_dam_w_2006} on the Keck-II telescope on May 25, 2023 (UT), approximately three years after the peak of the event.
In our Keck images (Figure 1c), we measure a total brightness of $\ks$=$16.99\pm0.03$ at the event location within a circular aperture of radius 0.2\( ^{\prime\prime} \), which closely matches the infrared source brightness ranging from $\ks$=$16.95\pm0.06$ to $\ks$=$17.08\pm0.06$ for the four degenerate solutions (see methods).
Our high-angular-resolution imaging reveals that the blended light in OGLE-III pre-event imaging arose primarily from field stars within 0.5\( ^{\prime\prime} \) to the west and north-west directions (Figure 1b/c).
As shown in Table 1, our aperture photometry constrains any excess flux above the source flux to be at least around two magnitudes fainter (at the 3-sigma level) than the expected brightness of the lens star if it were on the main sequence.
Therefore, we reject the MS hypothesis and conclude that the primary lens star, i.e. the planet host, must be a WD.

The conclusion that the primary lens is a WD calls for a re-examination of the four degenerate light-curve models. We find that the two south-east (SE) solutions are unlikely as both of them would require an extremely-low-mass (ELM) WD below 0.3 M$_\odot$.
ELM WDs (e.g.\ \cite{brown_elm_2020,brown_elm_2022}) are a rare class of WDs formed exclusively through binary interactions, where the companion star strips away the stellar envelope from the ELM WD progenitor via either common envelope evolution or stable mass transfer, before the progenitor star could initiate helium burning (e.g.\ \cite{sun_formation_2018,li_formation_2019}).
We can immediately rule out the existence of such massive companions to the lens star, as the light-curve models constrain the total lens mass as opposed to the primary lens mass in the case of close-in binaries.
It is also difficult to attribute the formation of an ELM WD to the close-in brown dwarf companion under the close-SE model, as binary evolutionary models \cite{zorotovic_post-common-envelope_2010,belloni_formation_2024} predict that a brown dwarf companion could only eject the envelope of the WD progenitor if it spiraled in to a much closer ($\lesssim 0.01$\,au) orbit or first interacted with the progenitor when it was an AGB star, when the core mass have grown to more than $\sim0.5M_\odot$.

On the other hand, the two NE models do not require a compact binary formation (ELM WD) interpretation.
As the finite age of the universe limits the lowest mass WD to form via the single star evolution,
we impose a host-mass lower limit of $M>0.45M_\odot$ based on WD population statistics (e.g.\ \cite{falcon_gravitational_2010,kilic_100_2020}), which serves as a Bayesian prior that refines the lens system properties.
Under this additional constraint, both the close-NE and wide-NE solutions indicate an approximately 1.7--1.9$M_\oplus$ planet at a projected separation around 2.1 au, with a host mass near 0.5$M_\odot$ (Table 2).
The planet mass is consistent with a rocky composition, and the corresponding planet size would be merely 20\% greater than Earth's radius from mass-radius relationships (e.g.\ \cite{chen_probabilistic_2016,otegi_revisited_2020}). Furthermore, we infer from WD initial-final mass relations \cite{cummings_white_2018} that the progenitor (MS) mass is likely around 1--2 $M_\odot$.

We then infer the planet's physical separation from its projected separation using orbital motion effects \cite{dong_microlensing_2009,skowron_binary_2011} in the light-curve models (Extended Data Table 1; see Methods). We adopt a log-uniform prior on the physical separation, and model the planet orbit for different assumed eccentricities.
As illustrated in Figure 2, the posterior distribution for the physical separation is bimodal, which reflects two distinctly allowed orbital configurations (Extended Data Figure 1).
The planet is most likely near greatest elongation in a significantly inclined orbit, which implies that the physical separation is near the projected separation.
Alternatively, the planet is near conjunction on a nearly edge-on orbit, which implies a physical separation of $\gtrsim 10$ au.
The former scenario is substantially favored for eccentricities up to $e<0.2$, for which we may place an upper limit to the physical separation at 2.3 au with 80--90\% confidence.
Due to tidal circularization during the host-star red-giant phases (e.g. \cite{jackson_tidal_2008,jones_properties_2014}), we consider it reasonable to assume that the current planet orbit indeed has low eccentricity.

For $e<0.2$, the close-orbit case ($d\sim2.1$ au) is formally favored by a Bayes Factor of around 5--10, which only constitutes \textit{substantial} but not \textit{strong} evidence \cite{kass_bayes_1995}.
Therefore, the extent to which the wide-orbit case ($d\gtrsim 10$ au) may be ruled out is sensitive to the adopted physical separation prior, which is complicated by the fact that the population of terrestrial planets at such separations remains largely unexplored.
Canonical planet formation theory expects terrestrial planets to form predominantly within the water ice line at around 3 au for a sun-like star (e.g. \cite{ida_toward_2004}).
However, processes such as planet-planet gravitational interactions during early stages of planet formation could scatter low-mass planets to very wide separations or outright eject them \cite{rasio_dynamical_1996}.
Statistics from short time-scale ($t_{\rm E}\lesssim0.5$ day) microlensing events indicate that wide-orbit ($\gtrsim10$ au) and free-floating low-mass planets (FFP) combined are at least as abundant as the known population of close-orbit planets \cite{gould_free-floating_2022,sumi_free-floating_2023},
but current follow-up observations are insufficient to distinguish between the two scenarios \cite{mroz_free-floating_2023}.
Therefore, if a considerable fraction of such microlensing FFP candidates are confirmed to be bound planets in the future (via direct detection of host star), then it becomes more likely for KB200414Lb to have a wide orbit than currently inferred.

Similarly (but for a different reason), the brown dwarf companion takes on either a very close or very wide projected separation, which would indicate distinct evolutionary histories (Figure 3).
To end up in a close-in orbit of $\gtrsim$0.2 au under the close-NE model, the BD companion would likely have gone through a period of common envelope evolution with the WD progenitor and successfully ejected the stellar envelope.
However, most known post-common-envelope binaries (PCEBs) have orbits smaller than 0.01\,au \cite{nelson_minimum_2018}. Several WD binaries with MS companions are known with separations of order 0.2\,au that are suspected to be PCEBs \cite{kruse_koi-3278_2014,kawahara_discovery_2018,yamaguchi_wide_2024}. 
Models are only able to explain these wider PCEBs if mass transfer was first initiated during the AGB phase of the progenitor of the WD, when its envelope is expected to be loosely bound and little gravitational energy is required to unbind it \cite{izzard_post-agb_2018,yamaguchi_wide_2024,belloni_formation_2024}.
Under this scenario, the BD initial orbit around the MS host is expected at 3--6 au \cite{yamaguchi_wide_2024}.
Nevertheless, even if this CEE pathway remains valid for a substantially less massive BD, long-term orbital stability for the system (e.g. \cite{deck_first-order_2013}) would require the planet to be on an initially wide orbit ($d\gtrsim10$ au), which is already disfavored by the planet orbital model.

Given the combination of evidences against the close-NE model, we conclude that the wide-NE model (Scenario 2; Figure 3) is the most favored scenario, where both the planet and the brown dwarf avoided interacting with the WD progenitor.
In this case, this system may provide a possible glimpse into the distant future of our solar system.
While Venus will eventually be engulfed and Mars will most certainly survive, the final fate of the Earth is rather uncertain and critically depends on the stellar-mass-loss rate during the solar RGB phase \cite{guo_effects_2016}, which remains poorly constrained \cite{mcdonald_mass-loss_2015}.
Certain models predict that the Earth may be engulfed during the solar tip-RGB phase due to tidal interactions and dynamical drag \cite{schroder_distant_2008,lanza_residual_2023}.
Nevertheless, if Earth had indeed survived, then its orbit is expected to expand to around twice its current size, comparable to the current orbit for KB200414Lb.
Therefore, the Earth-mass planet KB200414Lb likely represents a similar yet more fortuitous future compared to our own planet Earth.

\bigskip
\newpage

\section*{Methods}

\subsection*{Observations} \label{sec:phot}
We observed the location of the planetary microlensing \cite{mao_gravitational_1991,gould_discovering_1992} event KB200414 \cite{zang_earth-mass_2021} using the wide mode of the NIRC2 camera on the Keck-II telescope on May 25, 2023 (UT) under program U152 (PI: J.S.\ Bloom; Science-PI: K.\ Zhang). The pixel scale is 0.04\( ^{\prime\prime} \)/pixel with a 40\( ^{\prime\prime} \) by 40\( ^{\prime\prime} \) field of view. Five deep images are taken with 30 seconds of exposure per image for relative photometry on the target. Two shallow images are taken each with 15 seconds of total integration time, which consist of fifteen co-adds of 0.5-second exposures. The shallow image has a brighter saturation limit and is used for calibration to the VVV photometric system. The shallow and deep images are non-linearity corrected \cite{metchev_palomarkeck_2009}, sky-subtracted, flat-fielded, and averaged into two master images.

We identify the target in the Keck image by transforming the magnified source location in the CFHT image (Figure 1b) into the Keck frame. A linear transformation between the two frames is derived using ten reference stars listed in Supplementary Table 1, resulting in a residual standard error of 22.6 mas. We unequivocally identify the Keck star located at (502.43, 559.02) as the event location, which has an nominal offset from the CFHT source location of $22.0 \pm 22.6$ mas, i.e., within one pixel in the Keck image.

We then perform aperture photometry with a radius of five pixels (0.2\( ^{\prime\prime} \)) on the two stacked images using the \textit{photutils} package \cite{bradley_astropyphotutils_2022}. Eleven relatively isolated stars in the shallow image with 12.5<K$_{s, VVV}$<15.5 are calibrated to VVV DR4 aperture photometry \cite{minniti_vista_2010}, which results in a zero-point uncertainty of 0.03 mag. We then calibrated the deep image to the shallow image, which results in a calibrated target brightness of $\ks$=$16.99\pm0.03$.
Given the lens-source relative proper motion of $\sim$8 mas/year (Table 2), we may expect the lens-source separation to be $\sim$24 mas at the time of the Keck observations, much smaller than the $\sim80$ mas Keck PSF. Therefore, the target flux includes the combined flux from the lens and source stars. We note that the OGLE blended light may be attributed to four stars within 0.5\( ^{\prime\prime} \) to the west and north-west directions, which has a total brightness of $\ks\simeq16.8$. This is comparable to the source star brightness (see section below), which is also the case for the OGLE I-band blend.

\subsection*{Flux Constraints}
The source-star brightness was only measured in the $V$ and $I$ bands and slightly differs across models.
Since the follow-up observations were performed in the $\ks$ band, we first convert the $I$-band source brightness to the $\ks$ band from its intrinsic $(I-\ks)$ color and reddening $E(I-\ks)$.
To derive the extinction and reddening, we construct a ($I-\ks$) vs $\ks$ color-magnitude diagram (CMD) by cross matching OGLE-III and VVV catalog stars located within 2\( ^{\prime} \) of the location of KB200414 (Supplementary Figure 1). The VVV photometry is calibrated to 2MASS.
We measure the centroid of the red giant clump as $(I-\ks, \ks)_{\rm cl} = (2.49 \pm 0.01, 13.06 \pm 0.02)$.
For the intrinsic centroid of the red giant clump, we adopt
$(I - \ks , \ks)_{\rm cl,0} = (1.46\pm0.04, 12.89\pm0.04)$
\cite{nataf_reddening_2013,nataf_interstellar_2016}, which implies 
$E(I-\ks)=1.03\pm0.04$ and
$A_{\ks} = 0.17\pm0.04$.
We also cross check the $\ks$ extinction in color space. Using the OGLE extinction calculator, we derive reddening $E(V-I)=0.972$ and $E(J-\ks)=0.316$ \cite{gonzalez_reddening_2012} towards the sight-line of KB200414. Adopting the extinction law of ref.\ \cite{nishiyama_interstellar_2009}, we have $A_{Ks}=0.528\cdot E(J-\ks)=0.17$, which is in agreement with the CMD analysis.

We then derive the intrinsic $(I-\ks)$ source color from its intrinsic $(V-I)$ color, which was reported as $(V-I)_{\rm S,0}=0.84\pm0.03$ in ref. \cite{zang_earth-mass_2021}. Using color-color relations \cite{bessell_jhklm_1988} and the zero-point offset from $\ks$ to standard $K$ of 0.04 mag \cite{carpenter_color_2001}, we derive $(I-\ks)_{\rm S,0}=1.06 \pm 0.04$ and thus $(I-\ks)_{\rm S}=(I-\ks)_{\rm S,0}+E(I-\ks)=2.09 \pm 0.06$, which is used to convert the $I$-band source brightness (see Table 4 of ref. \cite{zang_earth-mass_2021}) into the $\ks$ source brightness listed in Table 1.

We derive the expected $\ks$ brightness for hypothetical main-sequence lenses using the MESA \cite{paxton_modules_2011} Isochrones and Stellar Tracks (MIST; \cite{choi_mesa_2016,dotter_mesa_2016}). The apparent brightness depends on the mass, distance, age, metalicity, and extinction experienced by the lens star. To rule out all possible main-sequence lenses, we must consider stellar properties that leads to the faintest brightness.
Therefore, we adopt metal-rich ([Fe/H]=0.5) isochrones and consider the faintest possible brightness over 100 Myr and 10 Gyr of age.

The mass and distance of the primary lens star is derived from the angular Einstein radius and the microlensing parallax as constrained by the light-curve models and source star properties. We directly adopt the published light-curve models of \cite{zang_earth-mass_2021} in the form of raw MCMC chains.
We searched for additional degenerate models using a machine-learning algorithm \cite{zhang_real-time_2021,zhang_ubiquitous_2022}, which did not yield new solutions but recovered the existing ones.
Note that the lens properties originally reported in Table 5 of \cite{zang_earth-mass_2021} applied both a Galactic model and rejected parameter samples that would result in the MS lens being brighter than the blend flux of $I=18.9$.
Since we have rejected the hypothesis that the primary lens is a MS star, we simply adopt a uniform prior, which results in slightly different reported values.

The angular Einstein radius is defined as
\begin{equation}
    \theta_E=\sqrt{\kappa M_{\rm L} \pi_{\rm rel}},
\end{equation}
where $\pi_{\rm rel}=\pi_{\rm L}-\pi_{\rm S}$ is the lens-source relative parallax and
\begin{equation}
    \kappa=\dfrac{4G}{c^2{\rm au}}\simeq8.144\,{\rm mas}/M_\odot.
\end{equation}

The microlensing parallax is defined as the lens-source relative parallax in units of the angular Einstein radius
\begin{equation}
    \pi_{\rm E}=\dfrac{\pi_{\rm rel}}{\theta_{\rm E}}=\sqrt{\dfrac{\pi_{\rm rel}}{\kappa M_{\rm L}}}.
\end{equation}
Therefore, the lens mass is derived as $M_{\rm L}=\theta_{\rm E}/\kappa/\pi_{\rm E}$ whereas the lens parallax is $\pi_{\rm L}=\pi_{\rm rel}-\pi_{\rm S}=\pi_{\rm E}\cdot\theta_{\rm E}-\pi_{\rm S}$. 
For the source parallax, we adopt a source distance of $D_{\rm S}=8.0\pm0.8$ kpc, which is derived using the triaxial G2 Galactic Bulge model originally adapted in \cite{zhu_toward_2017} for microlensing population studies.

Following ref.\ \cite{yang_kmt-2021-blg-0171lb_2022}, we derive the $\ks$ extinction experienced by the lens star (regardless of MS/WD) as
\begin{equation}
    A_{\ks}(D_L)=\int^{D_L}_{0}a_{\ks}\times n_d(D)dD,
\end{equation}
where $n_d(D)$ is the dust density at $D$, and $a_{\ks}$ is the extinction in units of mag\,kpc$^{-3}$ dust. We adopt an exponential Galactic dust distribution model where, in cylindrical coordinates,
\begin{equation}
    n_d(D)\propto \exp\left(\dfrac{|z(D)|}{z_d}-\dfrac{R(D)}{R_d}\right),
\end{equation}
where 
\begin{equation}
    z(D)=z_\odot+D\sin b\simeq z_\odot+Db,
\end{equation}
\begin{equation}
    R(D)=\sqrt{(R_\odot-D\cos b \cos l)^2+(D\cos b\sin l)^2}\simeq |R_\odot-D|.
\end{equation}
The dust length scales are adopted as ($R_D$, $z_d$) = (3.2, 0.1) \cite{li_three-dimensional_2018} and the location of the Sun is adopted as ($R_\odot$, $z\odot$) = (8.3, 0.023) kpc \cite{gillessen_monitoring_2009,maiz-apellaniz_spatial_2001}. The extinction constant is derived as $a_{\ks}$ = 0.67 by considering $A_{\ks}(D_{\rm S}) = 0.17$ and $D_{\rm S}$ = 8 kpc. The minimum expected brightness for MS lenses consistent with the light-curve models is reported in Table 1, with $\ks$ lens extinctions in the range of 0.03--0.06 mag.

\subsection*{White Dwarf Properties}
The age of the universe limits the lowest mass white dwarf that could be formed via single star evolution. 
CO white dwarfs are known to have a mass distribution sharply centered around 0.59 $M_\odot$, which drops off quickly for lower masses with essentially no WD found with $M<0.45M_\odot$ except for ELM/Helium WDs \cite{kilic_100_2020}. We therefore impose a host-mass lower limit of $M>0.45M_\odot$ as a Bayesian prior to further refine properties of the planetary system.
As low mass ($\sim0.5M_\odot$) WDs are already strongly favored by the light-curve models, the inferred host mass is relatively insensitive to the specific WD mass prior adopted, so long as some form of prior is applied to reject the regime of $M<0.45M_\odot$ where singular WDs are extremely uncommon.

We derive the expected $\ks$ CO WD brightness for the two NE solutions using the isochrone for $0.54M_\odot$ DA WDs under the BaSTI stellar evolution model \cite{salaris_large_2010}. We consider the possible WD brightness under a uniform cooling age distribution over 0.1 Gyr to 10 Gyr. We apply the same extinction scheme as for main-sequence lenses, which results in an expected WD lens brightness of $\ks\sim24$. As such, it would be favorable to directly observe the WD lens at the first light of the thirty-meter-class telescopes (est.\ 2030), at which point it will be separated from the glare of the source star by around 80 mas. It may also be possible to detect the WD lens with JWST.

\subsection*{Orbital Model}

Here, we infer the planet's physical separation ($d$) and semi-major axis ($a$) from its projected separation ($s$) by leveraging the microlensing orbital motion effect, which was included in the light-curve models originally published by \cite{zang_earth-mass_2021}.
The microlensing orbital motion effect considers the projected separation ($s$) and the relative angle ($\alpha$) as changing linearly in time and is parametrized as ($\ds$, $\da$).
Since the planetary light-curve feature occurred during a short 7-day window, this linear parameterization is likely sufficient, which we later validate by examining how much ($\ds$, $\da$) is actually predicted to change during this time frame.

We convert the planet orbital motion parameters ($\ds$, $\da$) for the North-East models to physical units under the host-mass lower limit, which are approximately $\dot{\alpha}=0.3\pm0.1$ rad yr$^{-1}$ and $\dot{s}=0.0\pm0.1$ au yr$^{-1}$ (see Extended Data Table 1).
Note that \cite{zang_earth-mass_2021} only considered orbital motion for the planet, but not for the BD. They estimated that doing so would require an additional $\mathcal{O}(10^6)$ CPU hours for each degenerate model. Moreover, they suggested that BD orbital motion is not expected to make a pronounced impact on the light curve, as the light-curve anomaly associated with the BD is less than half a day in duration.

We consider an orbital model with six parameters: host mass ($M$), semi-major axis ($a$), eccentricity ($e$), inclination ($i$), argument of periapsis ($\omega$), and the reference phase ($\phi_0$), which is defined as the difference between the reference time ($t_0$) in the light-curve model and the time of periastron ($t_{\rm peri}$), and normalized to the orbital period ($P$): $\phi_0=(t_0-t_{\rm peri})/P$. This parametrization allows the orbital model to become invariant to the orbital period and host mass, which we use to scale the orbital model as a separate step.

The physical separation and semi-major axis are deterministically related to the projected separation and orbital elements ($e, i, \phi_0, \omega$) via $d=s/f(\theta)$ and $a=s/g(\theta)$, where $\theta$ is a shorthand for the aforementioned orbital elements.
We first transform samples from the projected separation posterior (Table 2) into the physical separation posterior without the orbital motion constraints.
To this end, we sample a dense grid of orbital elements from a uniform prior for $\omega$ and $\phi_0$, and a sine prior for $i$, which facilitates an isotropic prior on the orbital plane.
We sample distinct eccentricities over [0, 0.5] at a step size of 0.1.
We then evaluate a grid of transforming factors $f(\theta)$ and $g(\theta)$ from the grid of orbital elements using the exoplanet package \cite{exoplanet:joss}. Finally, we acquire ($M,s$) samples from the light-curve posterior (Table 2) and apply the grid of transforming factors to derive samples of the physical separation. We then apply the same procedure for the semi-major axis.

Formally, we have applied a change of variables, where the physical separation posterior is related to the projected separation posterior (from the light-curve model) via
\begin{equation}
    p(d,\theta)=p(s,\theta)\left|\dfrac{\partial s}{\partial d}\right|=p(s,\theta)\cdot f(\theta)=p(s,\theta)\cdot \dfrac{s}{d},
\end{equation}
where $p(s,\theta)$ is a shorthand for $p(s=d\cdot f(\theta),\theta)$.
From the above equation, we may interpret $p(d,\theta)$ as a posterior distribution, where $p(s,\theta)$ is the (partial) likelihood of the projected separation and the physical separation prior is given by $p(d)\simeq1/d$, namely a log-uniform distribution. We have verified the log-uniform prior numerically given its importance in interpreting the final results.

We may write the above intermediate posterior as $p(d,\theta|s)$, since it only accounts for the projected separation measurement, not the orbital motion measurements. Observe that the full (taking into account all of $s,\sa$) and intermediate posteriors follows the same joint distribution $p(d,\theta,\ssa)$,
\begin{equation}
    p(d,\theta|\ssa)\propto p(d,\theta,\ssa)\propto p(d,\theta|s)\cdot p(\sa|M,s,\theta).
\end{equation}
Therefore, we may convert samples from the intermediate posterior to the full posterior with an importance weight of $p(\sa|M,s,\theta)$, namely the partial likelihood of the orbital motion constraints.
The predicted orbital motion is derived using finite difference on the aforementioned orbital element grid, which requires knowledge of the orbital period.
The host-mass associated with the projected separation (the $M,s$ samples from Table 2) underlying each parameter combination is used to derive the orbital period via Kepler's third law.
Therefore, our approach natively accounts for the covariance between $M$ and $s$, which circumvents the difficulty that $s$ is an observable whereas $M$ is a model parameter.
Therefore, we expect this novel approach to be useful for microlensing follow-up analysis in the future.

We first validate the linear orbital motion assumption by examining the extent to which $(\sa)$ is predicted to change during the planetary light-curve feature. We found that they change by merely $\mathcal{O}(10^{-3})$ au yr$^{-1}$ and $\mathcal{O}(10^{-5})$ rad yr$^{-1}$, which implies that linear orbital motion is a sufficient parametrization.

As we discuss in the main text, the bi-modality of the physical separation represents two distinct regions of orbital space that are allowed under the orbital model.
In Extended Data Figure 1, we visualize the marginal likelihood $p(\sa|i,\phi_0)$ for the inclination and reference phase, under different eccentricities.
To ease interpretation and without substantial loss of generality for mildly eccentric orbits, we fix the argument of periapsis to $\omega=\pi/2$ such that periastron and apastron occur at conjunction.
We may see that the planet is either near greatest elongation in a significantly inclined orbit, or near conjunction on a nearly edge-on orbit, with the former substantially favored. 

To interpret the origins of this degeneracy (bi-modality), let us first observe that if the planet were on a circular, face-on orbit, then given $a\simeq2.1$ au and $M\simeq0.5M_\odot$, and we may expect a constant $\da\simeq1.5$ rad yr$^{-1}$ from Kepler's third law, which is much greater than the measured $\da\simeq0.3$ rad yr$^{-1}$. Therefore, the orbit must be substantially inclined. Furthermore, the measured $\ds$ is close to zero, which indicates that the planet is either near conjunction or longest elongation, which are the two locations where the projected separation remains stationary. If the planet were near conjunction, then its physical separation would greatly exceed the projected separation, which leads to much longer orbital period that serves to reduce $\da$. 
This also explains why the conjunction scenario is favored at apastron (Extended Data Figure 1), where the planet's angular velocity is also intrinsically smaller.

\newpage
\section*{Tables}

\begin{table*}[htbp]
    \renewcommand\arraystretch{1.25}
    \centering
    \caption{\textbf{Properties of the lens system KMT-2020-BLG-0414L(bc) under a four-fold light-curve degeneracy, with a uniform Bayesian prior.} North-east and south-east indicate the direction of lens-source relative proper motion, which corresponds to the $u_0>0$ and $u_0<0$ solutions under the ecliptic degeneracy \cite{jiang_ogle-2003-blg-238_2004,poindexter_systematic_2005}. Close and wide relates to the projected separation of the brown dwarf. Reported values are median values with 68\% confidence intervals. Minimum MS lens brightness is defined as the minimum brightness for metal-rich ([Fe/H]=0.5) stars over 100 Myr to 10 Gyr of age. Microlensing parallax is the ratio between the lens-source relative parallax and the angular Einstein radius. 3-$\sigma$ excess flux is the upper limit to the excess flux at 99.7\% (3-$\sigma$) confidence, defined as the difference between the observed flux ($K_{\rm s}=16.99\pm0.03$) and the source flux.
    }
    \begin{tabular}{|l|l|l|l|l|l|}
    \hline
    Parameter & Unit & \multicolumn{2}{c|}{North-East} & \multicolumn{2}{c|}{South-East} \\
    \hline
    &&Close&Wide&Close&Wide\\
    \hline
    Primary lens mass&M$_\odot$&{0.45}$^{+0.10}_{-0.08}$&{0.36}$^{+0.08}_{-0.06}$&{0.25}$^{+0.06}_{-0.03}$&{0.19}$^{+0.06}_{-0.04}$\\
    Distance&kpc&{1.12}$^{+0.23}_{-0.19}$&{0.99}$^{+0.20}_{-0.17}$&{0.73}$^{+0.15}_{-0.09}$&{0.57}$^{+0.14}_{-0.09}$\\
    Minimum MS lens brightness ($\ks$)&mag&{16.32}$^{+0.27}_{-0.26}$&{16.84}$^{+0.24}_{-0.34}$&{17.16}$^{+0.16}_{-0.18}$&{17.25}$^{+0.16}_{-0.19}$\\
    \hline
    Einstein radius&mas&{1.73}$^{+0.06}_{-0.06}$&{1.65}$^{+0.07}_{-0.06}$&{1.62}$^{+0.05}_{-0.04}$&{1.64}$^{+0.05}_{-0.04}$\\
    Microlensing parallax&--&{0.45}$^{+0.10}_{-0.08}$&{0.54}$^{+0.12}_{-0.10}$&{0.76}$^{+0.12}_{-0.15}$&{0.99}$^{+0.21}_{-0.22}$\\
    \hline
    Source brightness ($\ks$)&mag&$17.08\pm0.06$&$16.99\pm0.06$&$17.03\pm0.06$&$16.95\pm0.06$\\
    3-$\sigma$ Excess brightness ($\ks$)&mag&18.63&19.06&18.88&19.34\\
    \hline
    \end{tabular}
    \label{tab:parm1}
\end{table*}

\begin{table*}[htbp]
    \renewcommand\arraystretch{1.25}
    \centering
    \caption{\textbf{Refined properties of the WD planetary system KMT-2020-BLG-0414L under a host-mass limit of $M>0.45M_\odot$.} Only the North-East (NE) models are shown as we have ruled out the South-East models. Lens-source relative proper motion measured north ($0^\circ$) to east ($90^\circ$) and in the heliocentric frame. The projected and physical separations are defined at the time of the event. Planet minimum physical separation for Scenario 1 arise from stability requirements during the host-star MS phase.}
    \begin{tabular}{|l|l|l|l|}
    \hline
    Parameter & Unit & Close-NE  & Wide-NE\\
    \hline
    WD mass&M$_\odot$&{0.51}$^{+0.08}_{-0.05}$&{0.49}$^{+0.06}_{-0.03}$\\
    Distance&kpc&{1.27}$^{+0.19}_{-0.12}$&{1.33}$^{+0.16}_{-0.12}$\\
    \hline
    WD flux ($\ks$)&mag&{23.8}$^{+0.7}_{-0.6}$&{24.0}$^{+0.6}_{-0.8}$\\
    Proper motion&mas yr$^{-1}$&{7.74}$^{+1.52}_{-0.63}$&{7.77}$^{+1.29}_{-0.76}$\\
    Proper motion direction&deg&{62.9}$^{+21.4}_{-21.7}$&{68.5}$^{+15.4}_{-20.2}$\\
    \hline
    Planet mass&M$_\oplus$&{1.75}$^{+0.31}_{-0.18}$&{1.87}$^{+0.27}_{-0.16}$\\
    Planet projected separation&au&{2.17}$^{+0.30}_{-0.17}$&{2.07}$^{+0.22}_{-0.11}$\\
    Planet physical separation ($e<0.2$)&au&$\gtrsim$10&{2.07}$^{+0.24}_{-0.09}$\\
    \hline
    BD mass&M$_{\rm J}$&{32.4}$^{+4.8}_{-2.6}$&{27.0}$^{+4.0}_{-3.1}$\\
    BD projected separation&au&{0.20}$^{+0.03}_{-0.02}$&{22.3}$^{+2.4}_{-1.5}$\\
    \hline
    \end{tabular}
    \label{tab:parm1}
\end{table*}

\setcounter{table}{0}
\renewcommand{\tablename}{Extended Data Table}

\begin{table*}[htbp]
    \renewcommand\arraystretch{1.25}
    \centering
    \caption{\textbf{Orbital motion parameters for the NE models.} Converted to physical units under the WD mass prior ($M>0.45M_\odot$) and shown as the mean values and standard deviations of the respective posterior distribution.}
    \begin{tabular}{|c|c|c|}
    \hline
    &Close-NE&Wide-NE\\
    \hline
    $\dot{\alpha}$ (rad yr$^{-1}$)&$0.30\pm0.15$&$0.27\pm0.11$\\
    $\dot{s}$ (au yr$^{-1}$)&$0.00\pm0.13$&$-0.08\pm0.15$\\
    \hline
    \end{tabular}
    \label{tab:parm1}
\end{table*}
\newpage

\setcounter{table}{0}
\renewcommand{\tablename}{Supplementary Table}

\begin{table*}[htbp]
    \renewcommand\arraystretch{1.25}
    \centering
    \caption{\textbf{Ten stars used to derive the astrometric transformation between the Keck and CFHT images.} The target location is listed at the bottom of the table.}
    \begin{tabular}{|cc|cc|}
    \hline
    \multicolumn{2}{|c|}{Keck}&\multicolumn{2}{c|}{CFHT}\\
    \hline
    x [pix]&y [pix]&x [pix]&y [pix]\\
    \hline
    433.52&247.47&336.25&276.74\\
    259.35&269.19&298.84&281.24\\
    634.16&319.38&379.09&292.43\\
    135.42&477.38&271.93&326.32\\
    520.32&512.34&354.35&334.07\\
    851.79&579.82&425.37&348.92\\
    248.57&639.97&296.00&361.34\\
    272.59&701.63&301.06&374.74\\
    445.83&713.47&338.22&377.35\\
    279.30&880.33&302.52&413.54\\
    \hline
    502.43&559.02&350.53&344.01\\
    \hline
    \end{tabular}
    \label{tab:parm1}
\end{table*}

\newpage

\section*{Figures}

\begin{figure}[htbp]
    \centering
    \includegraphics[width=\columnwidth]{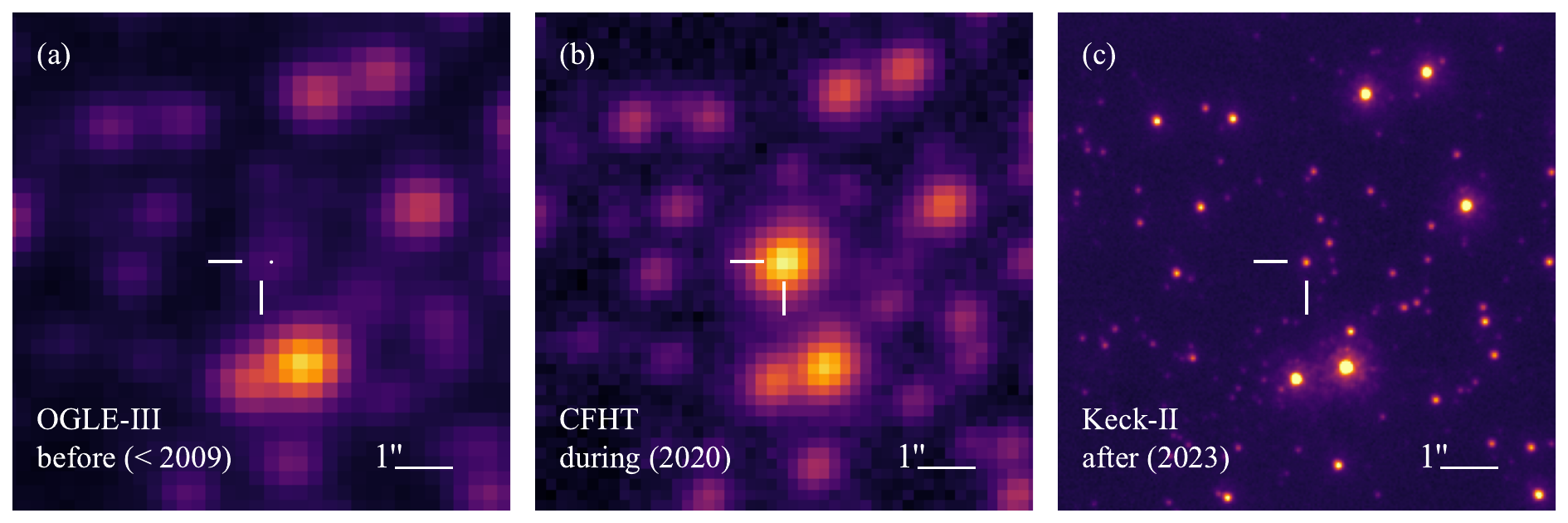}
    \caption{\textbf{OGLE-III, CFHT, and Keck-II imaging of KMT-2020-BLG-0414 taken before, during, and after the event. a}, OGLE-III $I$-band image taken over 2002--2009. The event location is centered on the cross-hair. The OGLE-III baseline (catalog) object is 0.18\( ^{\prime\prime} \) west and 0.01\( ^{\prime\prime} \) south of the event location, as indicated by the white dot. \textbf{b}, CFHT/MegaCam $i$-band image taken 2.2 days after the peak of the event. \textbf{c}, Keck AO $\ks$-band imaging reveals that the blend flux associated with the OGLE-III baseline object is predominantly attributed to field stars to the west/north-west within 0.5\( ^{\prime\prime} \).}
    \label{fig:image}
\end{figure}

\newpage

\begin{figure}[htbp]
    \centering
    \includegraphics[width=0.7\columnwidth]{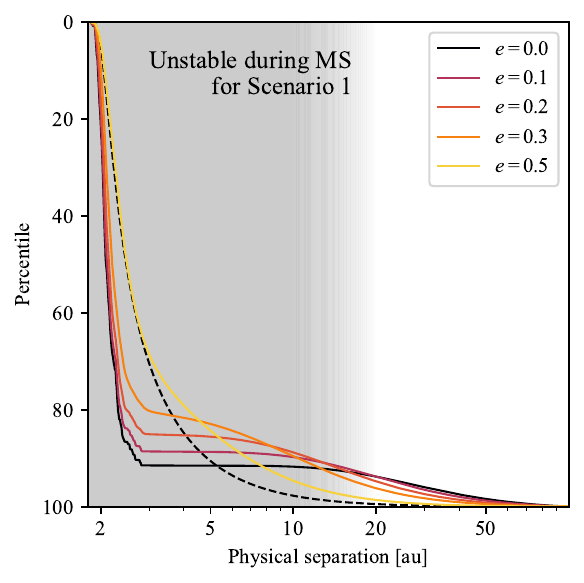}
    \caption{\textbf{The planet's physical separation from the WD host during the peak of the event as inferred from its projected separation and the microlensing orbital motion effect}. The cumulative distribution function (CDF) for the marginal posterior distribution of the physical separation under a log-uniform prior is shown for different eccentricities. The difference between the Close-NE and Wide-NE solutions is minimal (see Supplementary Figures 2 \& 3) and their mean value is displayed. The CDF for the physical separation distribution without the orbital motion constraints is shown for $e=0$ in the dashed line for comparison. The shaded region indicates the planetary orbits that would have been initially unstable with the BD orbit during the host-star MS phase.}
    \label{fig:image}
\end{figure}

\begin{figure}[htbp]
    \centering
    \includegraphics[width=\columnwidth]{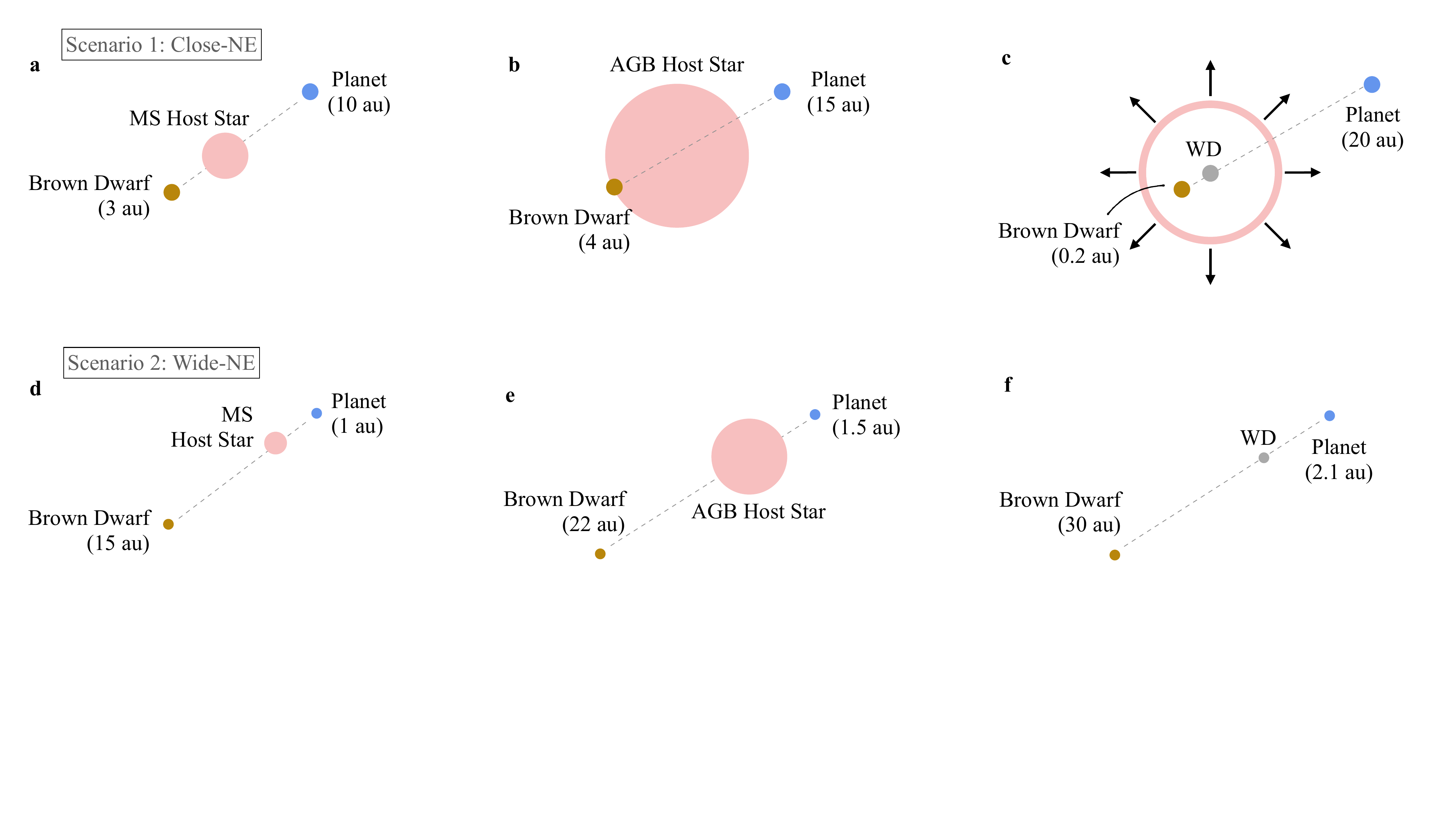}
    \caption{\textbf{Illustration of possible system evolutionary histories under the Close-NE (a--c) and Wide-NE (d--f) models.} Objects and orbits are not drawn to scale. Separations are representative values. \textbf{(a)} Initial configuration: close-in brown dwarf and wide-orbit planet. \textbf{(b)} Orbits expand due to host-star mass loss. AGB host star overflows Roche Lobe. Brown dwarf enters common envelope. \textbf{(c)} Common envelope ejected as brown dwarf orbit reduces to 0.2 au. Planetary orbit further expands. \textbf{(d)} Initial configuration: close-in planet and wide-orbit brown dwarf. \textbf{(e)} Orbits expand due to host-star mass loss. Brown dwarf and planet avoid interacting with the AGB host star. \textbf{(f)} Orbits continue to expand.}
    \label{fig:image}
\end{figure}

\setcounter{figure}{0}
\renewcommand{\figurename}{Extended Data Figure}

\begin{figure}[htbp]
    \centering
    \includegraphics[width=\columnwidth]{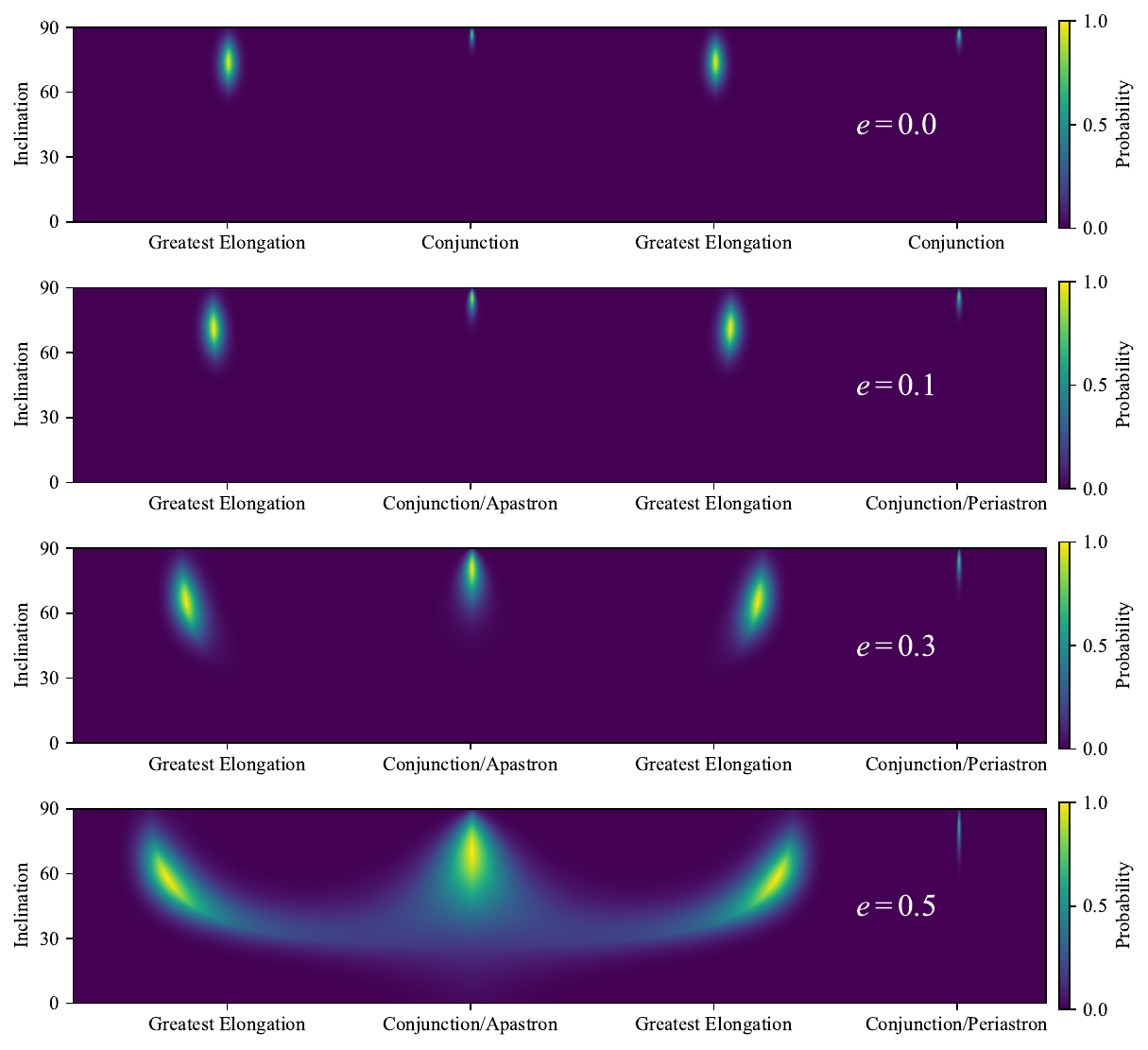}
    \caption{\textbf{Marginal likelihood for the inclination and orbital phase for different assumed eccentricities.} Shown for the special case of $\omega=\pm\pi/2$ where apastron and periastron are aligned with conjunction for eccentric orbits.}
    \label{fig:image}
\end{figure}

\setcounter{figure}{0}
\renewcommand{\figurename}{Supplementary Figure}

\begin{figure}[htbp]
    \centering
    \includegraphics[width=0.8\columnwidth]{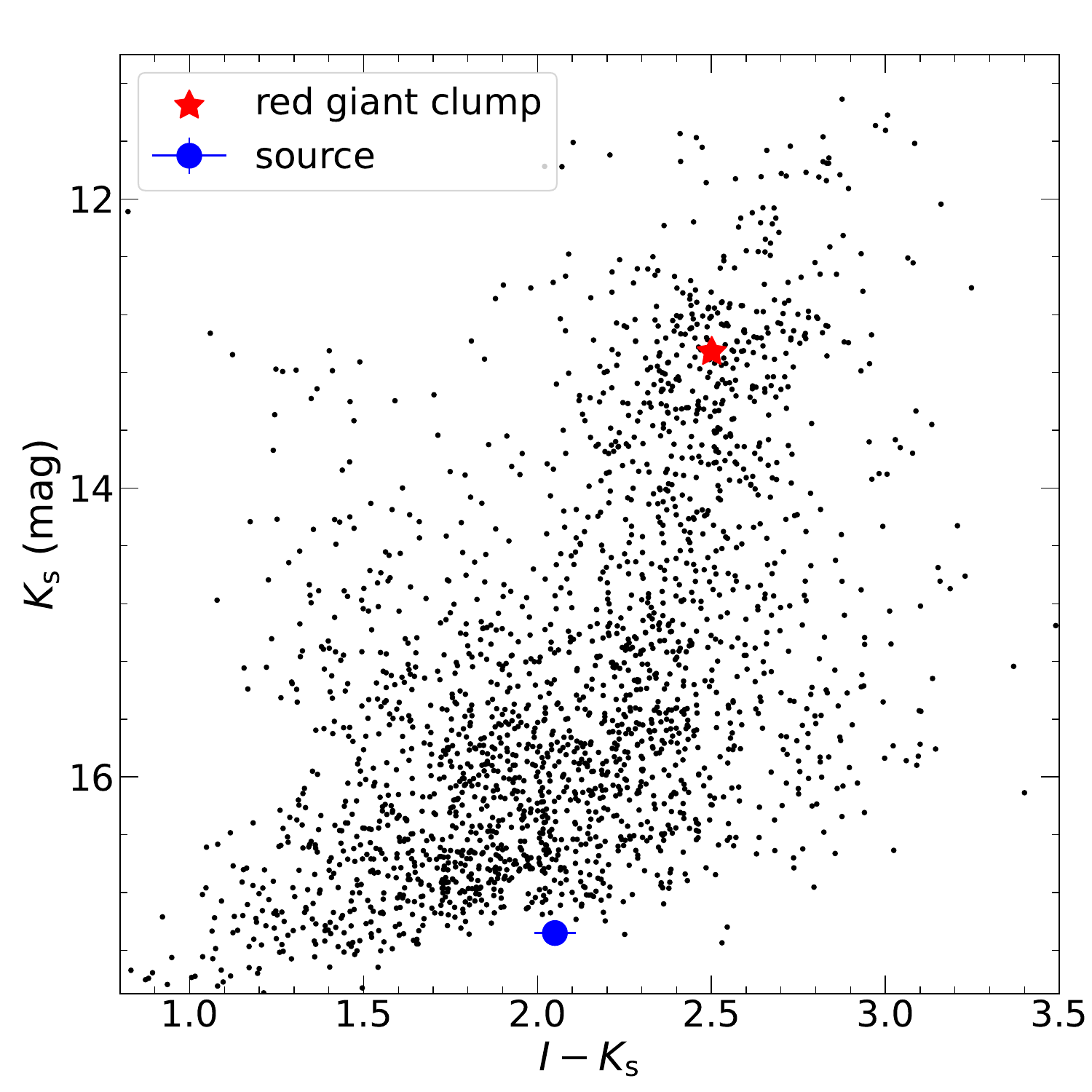}
    \caption{\textbf{OGLE-$I$ and 2MASS-$\ks$ color magnitude diagram for a 240\( ^{\prime\prime} \) $\times$ 240\( ^{\prime\prime} \) box centered on the location of KB200414.} The red asterisk and the blue dot represent the centroid of the red-giant clump and the source star, respectively.}
    \label{fig:cmd}
\end{figure}

\begin{figure}[htbp]
    \centering
    \includegraphics[width=\columnwidth]{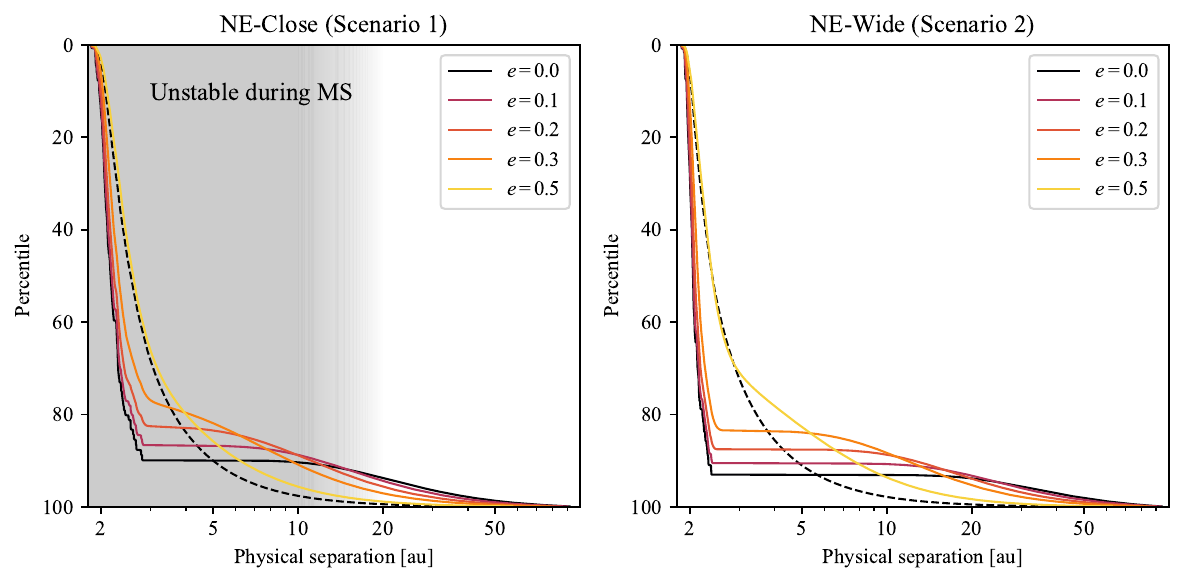}
    \caption{\textbf{The physical separation of the planet during the peak of the event as inferred from the projected separation and the microlensing orbital motion effect}. Identical to Figure 2 but displayed separately for the Close-NE and Wide-NE solutions. The cumulative distribution function (CDF) for the marginal posterior distribution of the physical separation under a log-uniform prior is shown for different eccentricities. The CDF for the physical separation distribution without the orbital motion constraints is shown for $e=0$ in the dashed line for comparison. The shaded region indicates the planetary orbits that would have been initially unstable with the BD orbit during the host-star MS phase.}
    \label{fig:image}
\end{figure}

\begin{figure}[htbp]
    \centering
    \includegraphics[width=\columnwidth]{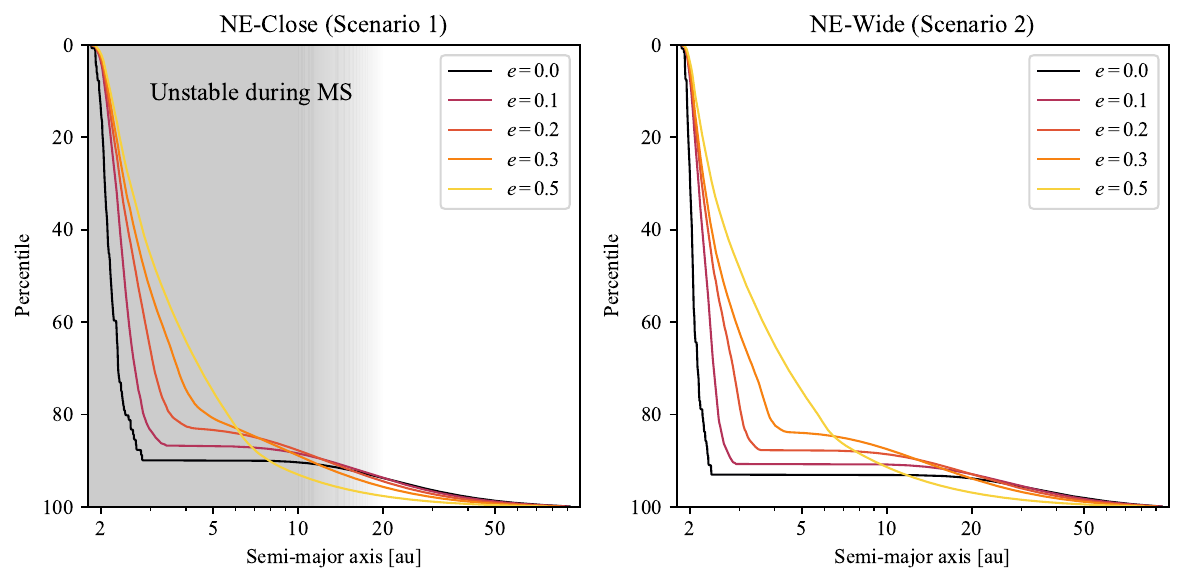}
    \caption{\textbf{The planet semi-major axis as inferred from the projected separation and the microlensing orbital motion effect}. The cumulative distribution function (CDF) for the marginal posterior distribution of the semi-major axis under a log-uniform prior is shown for different eccentricities. The shaded region indicates the planetary orbits that would have been initially unstable with the BD orbit during the host-star MS phase.}
    \label{fig:image}
\end{figure}

\newpage
\bigskip

\newpage

\section*{Data Availability Statement}
The reduced Keck images are available at https://zenodo.org/records/13128167. The raw data will be available on the Keck Observatory Archive (https://koa.ipac.caltech.edu/) after the 18-month proprietary period.

\section*{Acknowledgments}
K.Z. is supported by the Eric and Wendy Schmidt AI in Science Postdoctoral Fellowship, a Schmidt Futures program.
K.Z. and J.S.B. were partially supported by the Gordon and Betty Moore Foundation and a grant from the National Science Foundation (award \#2206744).
W.Z. acknowledges the support from the Harvard-Smithsonian Center for Astrophysics through the CfA Fellowship.
W.Z. and S.M. acknowledge support by the National Natural Science Foundation of China (Grant No. 12133005).
J.R.L. acknowledges support from the National Science Foundation under grant No. 1909641 and the Heising-Simons Foundation under grant No. 2022-3542.
W.Z. thanks Hanyue Wang for fruitful discussions on the Keck proposal.
This research has made use of the KMTNet system operated by the Korea Astronomy and Space Science Institute (KASI) and the data were obtained at three host sites of CTIO in Chile, SAAO in South Africa, and SSO in Australia.
Data transfer from the host site to KASI was supported by the Korea Research Environment Open NETwork (KREONET).
Some of the data presented herein were obtained at Keck Observatory, which is a private 501(c)3 non-profit organization operated as a scientific partnership among the California Institute of Technology, the University of California, and the National Aeronautics and Space Administration. The Observatory was made possible by the generous financial support of the W.\ M.\ Keck Foundation. The authors wish to recognize and acknowledge the very significant cultural role and reverence that the summit of Mauna Kea has always had within the Native Hawaiian community. We are most fortunate to have the opportunity to conduct observations from this mountain.
This research uses data obtained through the Telescope Access Program (TAP), which has been funded by the TAP member institutes.
Partly based on observations obtained with MegaPrime/MegaCam, a joint project of CFHT and CEA/DAPNIA, at the Canada-France-Hawaii Telescope (CFHT) which is operated by the National Research Council (NRC) of Canada, the Institut National des Science de l'Univers of the Centre National de la Recherche Scientifique (CNRS) of France, and the University of Hawaii. The observations at the Canada-France-Hawaii Telescope were performed with care and respect from the summit of Maunakea which is a significant cultural and historic site.

\section*{Author Contributions}
K.Z. reduced the Keck data, developed the probabilistic framework for inferring the planet's physical separation, led the overall analysis and interpretation, and wrote the manuscript. K.Z., K.E.B, and E.A. developed the interpretation of the system evolutionary history. K.Z. and W.Z. conceived of the observations and led the writing of the Keck proposal. W.Z. contributed to the extinction and lens light analysis. K.Z. and J.S.B. obtained the observing time as the Science-PI and PI of Keck program U152. J.R.L., S.T., J.S.B, and N.L. contributed to observing. All co-authors participated in discussions and contributed to the revision of the manuscript.

\section*{Competing Interests}

We declare no competing interests.

\section*{Corresponding author}

Correspondence and requests for materials should be addressed to Keming Zhang\\ (kemingz@berkeley.edu) or Weicheng Zang (weicheng.zang@cfa.harvard.edu).

\end{document}